\documentstyle[12pt]{article}

\newcommand{\singlespace}{
  \renewcommand{\baselinestretch}{1}\large\normalsize}
\newcommand{\doublespace}{
  \renewcommand{\baselinestretch}{1.6}\large\normalsize}

\begin{document}

\hyphenation{mono-pole mono-poles lat-tice}

\begin{titlepage}

\begin{tabbing}
\` ILL-(TH)-93-\#15 \\
\` July, l993 \\
\end{tabbing}

%\vspace*{.7in}
\vspace*{1.0in}

\begin{center}
{\bf Large Loops of Magnetic Current and Confinement\\
in Four Dimensional $U(1)$ Lattice Gauge Theory \\}
\vspace*{.5in}
John D. Stack\\
\vspace*{.2in}
{\it Department of Physics \\
University of Illinois at Urbana-Champaign \\
1110 W. Green Street \\
Urbana, IL 61801 \\}
\vspace*{.2in}
and \\
\vspace*{.2in}
Roy J. Wensley \\
\vspace{.2in}
{\it Department of Mathematical Sciences \\
Saint Mary's College \\
Moraga, CA 94575 \\
}
\vspace*{0.3in}
{\it (submitted to Physical Review Letters)}
\end{center}

\begin{tabbing}
\` PACS Indices: 11.15.Ha\\
\` 12.38.Gc\\
\end{tabbing}

\hspace{2.0in}

\end{titlepage}
\vfill\eject

\doublespace
\pagestyle{empty}

\begin{center}
{\bf Abstract}
\end{center}

\noindent We calculate the heavy quark potential from the magnetic current
due to monopoles in
four dimensional $U(1)$ lattice gauge theory.  The magnetic current is
found
from link angle configurations
using the DeGrand-Toussaint identification method.  The link angle
configurations are
generated in a cosine action simulation
on a $24^4$ lattice.
The magnetic current is resolved into large loops which wrap around the
lattice
and simple loops which do not.  Wrapping loops are
found only in the confined phase.  It is shown that the long
range part of the heavy quark potential, in particular the string tension,
can be calculated solely from the large, wrapping loops of magnetic
current.

\vspace*{.5in}

\newpage
\pagestyle{plain}

      In this paper, we report new results on confinement via monopoles for
$U(1)$ lattice gauge theory
in four dimensions.  Our main result is that the confining part of the
heavy quark potential, in particular the string tension, is determined
solely by large loops of magnetic current.
It has been established for some time that large loops which extend over
the entire lattice are present
only in the confined phase of the theory~\cite{wrap1,gupta}.
Their presence can now be quantitatively tied to
the string tension.  Our work is carried out on a
$24^4$ lattice, near the deconfining transition.

	The role of monopoles in $U(1)$ lattice gauge theory
is seen most clearly
using the Villain~\cite{Villain} or periodic gaussian form of the $U(1)$
theory.
Under a dual transformation, the usual link angle description goes over
into
one involving an integer-valued  magnetic current
$m_{\mu}(x)$,
defined on the links of the dual lattice~\cite{Banks}.
 The link angle path integral
becomes a sum over all possible configurations of magnetic current.
In this monopole representation, the system can be visualized as a plasma
of
magnetic monopoles moving on Euclidean world lines,
interacting via photon exchange.

In either representation, a Wilson loop calculation is needed to determine
the heavy quark potential.
In the link angle representation, a Wilson loop is specified  by the
exponential of  a line
integral :
 \begin{equation}
W(R,T)=\left<\exp\left(i\sum_{x}
\theta_{\mu}(x)J_{\mu}(x)\right)\right>_\theta,
\label{eqnlang}
\end{equation}
where the
integer-valued electric current $J_\mu$ is non-vanishing on the
rectangular $R \times T$
loop contour,
and $<\cdot>_\theta$  denotes the expected value taken over configurations
of link angles $\theta_{\mu}(x)$.

In the monopole representation, the determination of a Wilson loop involves
the exponential of an area
integral over a surface with the loop contour as its
boundary~\cite{Banks,ws2}.
The electric current $J_{\mu}$ is first expressed as the curl of
a Dirac sheet variable~\cite{dirac};
where $\partial_{\nu}$ denotes a discrete derivative.
The sheet variable $D_{\mu\nu}$
is nonunique. For  $|J_{\mu}|=1$,   a specific choice is made by
setting
$D_{\mu\nu}=1$ on the
plaquettes of an (arbitrary)
open \mbox{surface} with boundary  $J_{\mu}$, and $D_{\mu\nu}=0$ on all other
plaquettes.  The area integral
represents the dual flux set up by the magnetic current through this
surface.
To compute it, we define the magnetic vector potential
\begin{equation} A_\mu(x)=
\sum_{y}v(x-y)m_{\mu}(y),
\label{eqnamag}
\end{equation}
where $v$ satisfies $-\partial\cdot\partial\,
v(x-y)=\delta_{x,y}$.  The
field strength
is given by
\mbox{$F_{\mu\nu}\equiv\partial_{\mu}A_{\nu}-\partial_{\nu}A_{\mu}$.}
In terms of
$D_{\mu\nu}$
and the dual field strength
$F^{*}_{\mu\nu}(x)=\frac{1}{2}\epsilon_{\mu\nu\alpha\beta}F_{\alpha\beta}(x)
$,
the monopole representation of a Wilson loop is finally given by:
\begin{equation}
W(R,T)=W_{ph}(R,T)\cdot
                \left<\exp\left(\frac{i2\pi}{2}\sum_{x}
D_{\mu\nu}(x)
                   F^{*}_{\mu\nu}(x)\right)\right>_m ,
\label{eqnfac}
\end{equation}
where
$<\cdot>_m$ denotes the sum over configurations of magnetic current.
The factor $2\pi$ which appears in the exponent of Eq.~(\ref{eqnfac})
arises
from the Dirac condition on the product of electric and  magnetic charge,
and guarantees that the value of a Wilson loop is independent of the
surface
chosen to define  $D_{\mu\nu}$.  The prefactor in Eq.~(\ref{eqnfac})
describes one photon exchange between the quark and anti-quark:
\begin{equation}
W_{ph}(R,T)=\exp\left(-{e^2\over 2}\sum_{x,y}J_{\mu}(x)v(x-y)J_{\mu}(y)
\right),
\label{eqn_ph}
\end{equation}
where the electric coupling $e^2$  is related to
the coupling $\beta_{V}$ in the Villain action by
$e^{2}=1/\beta_{V}$.  The factor $W_{ph}(R,T)$ contributes a purely
perturbative Coulomb term to the heavy quark potential.

Fortunately,  the numerical evaluation of
Wilson loops via Eq.~(\ref{eqnfac}) does not
require a direct simulation in terms of the magnetic current
$m_{\mu}(x)$.
This is impractical owing to the long-range
interactions generated by photon exchange
between the monopole currents.
DeGrand and Toussaint~\cite{degrand} showed how to locate
monopoles directly in configurations of link angles.  In their procedure,
the plaquette angle
$\theta_{\mu\nu}(x)=\partial_{\mu}\theta_{\nu}-\partial_{\nu}\theta_{\mu}$
is resolved into a fluctuating
part $\bar{\theta}_{\mu\nu}(x)$, and an integer-valued
Dirac sheet variable $m^{*}_{\mu\nu}(x)$:
\begin{equation}
\theta_{\mu\nu}(x)=\bar{\theta}_{\mu\nu}(x)+2\pi m^{*}_{\mu\nu}(x),
\end{equation}
where $m^{*}_{\mu\nu}(x)
=\frac{1}{2}\epsilon_{\mu\nu\alpha\beta}m_{\alpha\beta}(x)$ and
$\bar{\theta}_{\mu\nu}(x)\in (-\pi,\pi) $.
 The magnetic current is then
given  by \mbox{$m_{\mu}(x)=\partial_{\nu}m_{\mu\nu}(x)$}.
This procedure
allows only values of $m_\mu\in[\pm2,\pm1,0]$, whereas in principle all
integer values are allowed.
However, at values of the coupling
%change delete 'of interest'
%of interest
%end change
near the deconfining transition, the values
$m_{\mu}=\pm1$ are overwhelmingly dominant;
even  $m_{\mu}=\pm2$ occurs only a
small fraction of the time. Thus  negligible error is caused by omitting
higher values of $m_\mu$.

The monopole current
%change -eliminate following phrase
%identified using the DeGrand-Toussaint method
%end change
can be
%change-omit 'directly'
%directly
used to
calculate physical quantities,
as well as merely to count monopoles.  In our previous work in
four dimensions~\cite{ws2},
we used
%change -eliminate followoing phrase
%DeGrand-Toussaint
%end change
the current $m_{\mu}(x)$ to evaluate Wilson loops from  Eq.~(\ref{eqnfac}).
Similarly in three dimensions~\cite{ws}, we used the
%change eliminate DeGrand-
%DeGrand-Toussaint
%end change
monopole
density $m(x)$ to evaluate Wilson loops using the $d=3$ analog of
Eq.~(\ref{eqnfac}).  In both cases the
resulting  heavy quark potential agreed with that
extracted from the link angles and
Eq.~(\ref{eqnlang}), to within statistical errors.  In the present work
on a $24^4$ lattice, we have again checked that potentials deduced directly
from link angles and Eq.~(\ref{eqnlang}) agree with those obtained from
the magnetic current and Eq.~(\ref{eqnfac}).
%change-insert following sentence
These calculations show that quantitative results on confinement can be
obtained
using topological objects.

	The derivation of  Eq.~(\ref{eqnfac}) as an exact formula is
only possible for Villain's form of the $U(1)$ theory. On the other
hand,
	Wilson's cosine form~\cite{wilson}
of the $U(1)$ action can be simulated much more efficiently.
   In our previous work~\cite{ws2,ws},
we have shown that Villain action results can be extracted from a cosine
action
simulation, if a simple coupling constant mapping is used.
More precisely, a simulation using the
cosine action at coupling $\beta_{W}$ is equivalent to a Villain action
simulation at coupling $\beta_{V}$, with $\beta_{V}$ related to
$\beta_{W}$ by~\cite{Villain,janke}
\begin{equation}
1/\beta_V=-2\ln\left({I_0(\beta_W) \over I_1(\beta_W)}\right),
\label{eqnmap}
\end{equation}
where $I_0$ and $I_1$ are modified Bessel functions.
Eq.~(\ref{eqnmap}) determines the
value of $1/\beta_V$, and hence $e^2$, which result from a
cosine action simulation at a  given value of $\beta_W$.  The factor
$W_{ph}$ in Eq.~(\ref{eqnfac}) is then completely determined.
The magnetic
current is identified
% change- eliminate Degrand-...
from the cosine action
link angle configurations
and the result used to calculate the second factor of Eq.~(\ref{eqnfac}).
%change reword next sentence
%Wilson loops calculated  using the cosine action and Eq.~(\ref{eqnmap})
The Wilson loops calculated  in this manner
using the cosine action and Eq.~(\ref{eqnmap})
differ from pure Villain action results by a harmless perimeter term.
%end change
The
%change -insert 'R-dependent terms in'
R-dependent terms in the
potentials agree within statistical errors~\cite{ws,wth}.

      To summarize, our simulations use the cosine form of the $U(1)$
action,
identify $m_{\mu}(x)$ using the DeGrand-Toussaint method, and evaluate
Wilson loops in the monopole representation.
%change -reword next sentence
%A heatbath algorithm was used to update the
%link angles~\cite{creutz}.
The link angle configurations were generated using a heatbath
algorithm~\cite{creutz}.
%end change
The calculation of $A_{\mu}(x)$
from $m_{\mu}(x)$ in Eq.~(\ref{eqnamag}) was
done using a four dimensional vectorized FFT~\cite{nobile,nobile2}.
The $R\times T$ rectangle lying in the Wilson loop plane was used as the
defining surface for $D_{\mu\nu}$.  Magnetic current configurations were
saved every $10$ sweeps.  After Wilson loops were obtained from these
configurations using Eq.~(\ref{eqnfac}), potentials were extracted
using standard methods.
The heavy quark potential
$V(R)$ was obtained from a straight line fit of
 $\ln W(R,T)$
 vs. $T$, over
an interval $T_{min}(R)$ to $T_{max}$, where $T_{min}(R)=R+2$ for
$R=2,3$, and $R+1$ otherwise, while $T_{max}=16$.  To determine the
string tension $\sigma$ and Coulomb coupling $\alpha$, the potentials
were then fitted to a linear plus Coulomb form,
$V(R)=\sigma R - \alpha /R+V_0$, over the interval $R=2$ to $R=7$.
The large string tension present in $U(1)$ makes it difficult to work at
larger values of $R$.
 Errors  in
physical quantities were
estimated  using both the jacknife method and binning the data into bins of
various size.

It is  well established for $U(1)$ that appreciable correlation lengths
occur only
in the immediate vicinity of the deconfining phase transition.
The location of
the phase transition moves to larger values of $\beta_W$ as the lattice
size increases, in a manner roughly consistent with finite size scaling
theory~\cite{gupta,azcoiti}.   Since only lattices of size up to $16^4$
were available in the published literature when we began our work,
it was first necessary for us to
locate the transition for a $24^4$ lattice.
To do this, we performed
a series of runs with various initial configurations
for $1.0100<\beta_W < 1.0120$, and monitored the
value of the $1 \times 1$ Wilson loop, $W(1,1)$.  For
$\beta_W \geq 1.0114$, the system always
reached  a state with $W(1,1)\sim 0.65$.  For
$\beta_W \leq 1.0112$, the system always
reached a state with $W(1,1)\sim 0.63$.  Subsequent analysis of the
heavy quark potential showed these two states to be deconfined and
confined, respectively.
%change reword next sentence
%Thus, our results locate the phase transition in
%the interval $1.0112 \leq  \beta_W < 1.0114$ for a $24^4$ lattice.
While we have not precisely located the deconfining phase transition,
consistency with our results requires that the transition be in the
interval $1.0112 \leq  \beta_W < 1.0114$ for a $24^4$ lattice.
%end change

To avoid problems associated with long autocorrelation times that occur
near the transition, we chose to use a run of 20,000 sweeps at
$\beta_W=1.0103$
for the results to be presented below.  At this value of $\beta_W$, the
correlation length is large enough to
observe the beginnings of continuum behavior,
but small enough to avoid problems
with long autocorrelation times.  The autocorrelation time $\tau$ measured
from the
$1\times 1$ Wilson loop was approximately 100 sweeps.
In Fig.(1), we show the heavy quark potential determined from
Eq.~(\ref{eqnfac}) for $\beta_W=1.0103$ using
 $936$ configurations of magnetic
current.  A linear-plus-Coulomb fit gave a string tension
of $\sigma = 0.56(1)$, and a Coulomb coupling
of $\alpha=0.30(2)$.  The total number of links carrying magnetic current
at this $\beta_W$ was
$98,400(800)$.  For comparison, we also show in Fig.(1), the potential
determined from Eq.~(\ref{eqnfac}) for the deconfined $\beta_W$ value,
$\beta_W=1.0114$, where 400 configurations of magnetic current were
analyzed.
At this value of $\beta_W$, the string tension was
statistically zero, while the Coulomb coupling was $\alpha = 0.22(4)$.
The total number of links carrying
%change insert 'magnetic' in front of current in next sentence
magnetic current was $40,000(300)$.
%end change

	We now turn to the resolution of the magnetic current into loops.
For every other configuration of magnetic current, or every 20 lattice
sweeps,
magnetic current loops were individually
identified and catalogued.
The loop-finding algorithm proceeded
by choosing a non-zero current link $m_\mu (x_0)$
and following
the current it carried through the lattice until a loop was completed
by a return
to the site
$x_0$.
This process was carried out repeatedly from different starting points
and ended when the entire
configuration of current  had been resolved into loops, with each
current-carrying link
belonging to a specific loop.
  The algorithm was deterministic:
when looking for an outgoing current link at a particular
lattice site, the direction
$\mu=1$ was chosen first, followed by $\mu=2,3,4$ .
  Intersections of loops did occur (i.~e. more
%change write 'one' not '1' in next sentence
than one outgoing link associated with a site), so the
%end change
set of loops  identified was not unique.
However, self-intersections of loops were
relatively rare, occurring with approximately
the same probability as
self-intersections of a purely random walk in $d=4$~\cite{itzykson}.

Each loop identified as
described above  forms a closed path of links and automatically
satisfies current
conservation.  Two different topologies are possible,
depending on whether
tracing out a loop also
involves winding or wrapping around the lattice.  A lattice with periodic
boundary conditions is a hypertoroid, so topologically nontrivial loops
which
wrap around the lattice are permitted.
To distinguish the two possibilities, the net current was
measured for each loop:
$$\Lambda_\mu=\sum_{x \in {\rm loop}}m_\mu(x).$$
It is easy to show that only loops which wrap around the lattice have
a non-vanishing $\Lambda_\mu$, and that the components of  $\Lambda_\mu$
must be integer multiples of the lattice size along an axis;
$\Lambda_\mu=n_{\mu} \cdot N$, for a cubic lattice of size $N^4$.
%chang -reword next sentence
%The integers $n_{\mu}$ can be interpreted as net charges.  For
The integer $n_{\mu}$ is the net current of the loop in the $\mu$th
direction.
For example,
for a loop with $\Lambda_{4} \neq 0$, the sum of the charge
density $m_4$ over
%change replace 'every' by 'each'
each spacial cube
or  ``time slice" will equal the net charge $n_4$. Likewise for other
directions.
While
an individual loop can have a non-vanishing $\Lambda_\mu$,  a net current
cannot actually occur on a finite lattice, so the sum of $\Lambda_\mu$ over
all
loops vanishes identically.
 In our data, it was typical for a loop  with   non-vanishing $\Lambda_\mu$
to be wrapped around the lattice several times in more than one direction.
Values of $|n_\mu|$ up to $10$ were observed.  In what follows we will use
the terms ``wrapped" for loops with non-vanishing $\Lambda_\mu$
and ``simple" for loops with vanishing $\Lambda_\mu$.

    An indication that wrapped loops are crucial for confinement is that
they
are
present only in the confined state and never are observed to occur in the
deconfined state.  As mentioned earlier, there
are almost twice as many links carrying magnetic  current in the confined
phase.  When
the current is broken up into loops, it is found that this excess in the
confined phase consists of a small number of wrapped loops.
At $\beta=1.0103$, out of the total of
$98,400(800)$ links carrying magnetic current,
$51,000(400)$
are in wrapped loops, the remainder in simple loops.  The average number
of wrapped loops is only $4.6(1)$, so that the typical wrapped loop
contains
thousands of links.  In contrast the average total number of simple loops
is
$6,210(6)$, of which $3,507(4)$ are in the form of
elementary one-plaquette current loops composed of four links.
The number of
simple loops  with a given number of links
decreases rapidly as the number of links
increases. Over $90\%$ of the
links in simple loops are included in loops with $60$ links or less.

	Since wrapped loops occur only in the confined phase, it is
natural to ask if they can explain the long range, confining part of the
heavy quark potential.
To investigate this, we
computed the heavy quark potential again using Eq.~(\ref{eqnfac}), but for
each
configuration, including only
the magnetic current from wrapped loops.  The results are shown (omitting
the
%change replace 'contribution from' with 'factor' in next sentence
photon factor $W_{ph}$) in Fig.~(2).
%end change
A linear plus Coulomb fit to the resulting
potential gave a string tension $\sigma_w=0.56(2)$, and a Coulomb term
$\alpha_w=0.09(1)$.  The string tension is within statistical errors of the
value $0.58(2)$ found
earlier from the heavy quark potential calculated using the full magnetic
current.
Next, we
carried out a similar calculation using only the magnetic
current from the simple loops.  This produced the rather flat potential
also shown in Fig.(2).   A linear plus Coulomb fit to this potential gives
zero string tension within statistical errors ($\sigma_s=.0000(6)$), and a
Coulomb term
%change delete word 'with'
%with
%end change
$\alpha_{s}=0.06(1)$.  The result of these two fits gives strong
evidence
that in the
long distance region, there is a clean separation between the contributions
of the two classes of loops.  Only the large, wrapped loops containing
thousands
of links contribute to the confining part of the potential.  This has been
demonstrated here only within certain error bars, but it may well be an
exact
statement.

	In the fits described above for $\beta_W=1.0103$,
the wrapped loops required a
Coulomb term with coupling $\alpha_w=0.09(1)$, while the simple loops
required
a Coulomb term with coupling $\alpha_s=0.06(1)$.
In addition, there is a
Coulomb term coming from the $W_{ph}$ factor in Eq.~(\ref{eqnfac}).  Using
Eq.~(\ref{eqnmap}) to evaluate $\beta_V(1.0103)$, gives
$\alpha_{ph}=0.13$ as the Coulomb coupling arising from $W_{ph}$.
Simply adding the various terms ,
we obtain $\alpha_{ph}+\alpha_{s}+\alpha_w= 0.28(2)$, consistent with
our previous result of $0.30(2)$ obtained with the full magnetic current.

%change -reword paragraph below
%Taking account of the  results from both large and small
%distance regions suggests comparing the potential determined from the
%full magnetic current and shown in Fig.(1) with the sum of the
%potentials arising from $W_{ph}$, wrapped loops, and simple loops.  The
%result is shown in Fig.(3), where it is seen that the agreement is quite
%good.  Additivity of the potential from the various
%contributions would imply  that the contributions from wrapped and simple
%loops factor in the average over configurations.
%As a
%check on factorization, we performed a fit to the ``potential"
%extracted from the ratio of Wilson loops assuming factorization to
%Wilson loops calculated with the full magnetic current:
%\begin{equation}
%	\left< W_{w}(R,T) \right>_m \cdot
%\left< W_{s}(R,T) \right>_m /\left< W_{w}(R,T)  \cdot
%W_{s}(R,T) \right>_m.
%\end{equation}
%The string tension and Coulomb coupling resulting from this were both
%zero to within statistical errors.
%If as we have
%argued, the string tension
%arises solely from the wrapped loops, then factorization would be
%expected to hold for large $T$ and $R$, where an area law holds for
%Wilson loops.  It is unexpected that it should hold
%in the small $R$ region where wrapped and simple loops of magnetic current
%both are producing Coulomb terms. Nevertheless factorization of the
%contributions of simple and wrapped loops does work, to within the
%statistical accuracy of our results.

%change reword of above
The results on the string tension and Coulomb coupling
are consistent with additivity of the potential over the various
contributions.
In Fig.(3), we compare the
potential determined from the full magnetic current and $W_{ph}$
(shown previously in Fig.(1)), with the potential obtained by summing the
contributions from $W_{ph}$, wrapped loops, and simple loops, plus a constant.
A glance
at Fig.(3) shows that the agreement is quite good.  Additivity of the
potential over the various contributions would imply that the contributions
from wrapped and simple loops factor in the average over configurations.
To check factorization, we performed a fit to the ``potential"
extracted from the ratio of Wilson loops assuming factorization to
Wilson loops calculated with the full magnetic current:
\begin{equation}
        \left< W_{w}(R,T) \right>_m \cdot
\left< W_{s}(R,T) \right>_m /\left< W_{w}(R,T)  \cdot
W_{s}(R,T) \right>_m.
\end{equation}
The string tension and Coulomb coupling resulting from this were both
zero to within statistical errors.  This shows that factorization and
therefore additivity of the potential is consistent with our data.  This
%change replaced 'large R region' with 'large R' in next sentence
%change remove 'only' and insert 'predominantly' in next sentence
%change start following sentence with 'However'
is not surprising at large R, where the R-dependence comes
predominantly from the wrapped loops.  However, in the small R region,
%change replaced 'are producing' with 'produce' in next sentence
both wrapped and simple loops produce Coulomb terms, and additivity is
not expected  to hold as an exact statement.
Nevertheless it appears to be a good approximation and holds within
the accuracy of our data.
%end change

	We have shown that confinement in $U(1)$ comes about through a
particular component of the magnetic current, the large wrapping loops.
%change added 'the large wrapping loops' in above sentence
%change reword rest of this paragraph
The small, simple loops contribute only to the Coulombic part
of the potential.  While the latter is intuitively reasonable,
still lacking is a
physical picture of how the large, wrapping loops of current
disorder the vacuum and produce the string tension.  The fact that these
loops extend over the whole lattice suggest that there are low mass
(perhaps massless) magnetically charged
excitations present in the confined phase.  We plan to
report elsewhere on this question as well as how the magnetic current
screens itself.
%change add the following sentence
The results obtained in our work are likely to have an impact on
the monopole approach to confinement in non-Abelian gauge theories.
%end change

This work was supported in part by the National Science Foundation under
Grant No. NSF PHY 92-12547.  The calculations were carried out on the Cray
Y-MP system at the National Center for Supercomputing Applications at the
University of Illinois, supported in part by the National Science
Foundation
under Grant No. NSF PHY920026N. R.\ J.\ W.\ would like to acknowledge
support from the Faculty Development Funds of Saint Mary's College of
California, and support from Lawrence Livermore National Laboratory.

%bibwr.tex

%
%   BIBLIOGRAPHY FILE
%

\singlespace

\newpage
\begin{figure}[h]
\includegraphics{fig2.ps}
\vspace{6.5in}
\caption{The quark potential calculated using the magnetic current
configurations for
$\beta_W=1.0103$ (triangles) and $\beta_W=1.0114$ (squares). The solid
lines
are linear-plus-Coulomb fits to the potentials.}
\end{figure}

\newpage
\begin{figure}[h]
\includegraphics{fig3.ps}
\vspace{6.5in}
\caption{The potential calculated using only the wrapping monopole loops
(triangles) and using
only the simple monopole loops (squares). The photon contribution from
$W_{ph}$ has not
been included. }
\end{figure}

\newpage
\begin{figure}[h]
\includegraphics{fig4.ps}
\vspace{6.5in}
\caption{Comparison of the quark potential calculated using the full
magnetic current and
photons (triangles) with the potential obtained by summing the potentials
determined separately from photons, wrapped loops, and simple loops
(squares).  The solid line is the linear-plus-Coulomb fit from Fig.~(1) for
$\beta_W=1.0103$.}
\end{figure}
\end{document}